\documentclass[12pt]{article}

\pdfoutput=1

\usepackage{mathtools}
\usepackage{color}
\usepackage{amsmath}
\usepackage{graphicx}
\usepackage{amssymb}
\usepackage{amsthm}

\numberwithin{equation}{section}

\title{Continuous representation for shell models of turbulence}

\author{Alexei A. Mailybaev\footnote{Instituto 
Nacional de Matem\'atica Pura e Aplicada -- IMPA, 
Est. Dona Castorina 110, 22460-320 Rio de Janeiro, RJ, Brazil. 
E-mail: alexei@impa.br.} 
}

\date{}

\begin{document}

\maketitle

\begin{abstract}
In this work we construct and analyze continuous hydrodynamic models in one space dimension, 
which are induced by shell models of turbulence. After Fourier transformation, 
such continuous models split into an infinite number of uncoupled subsystems, which are all 
identical to the same shell model. The two shell models, which allow such a construction, are considered: the dyadic 
(Desnyansky--Novikov) model with the intershell ratio $\lambda = 2^{3/2}$ and the 
Sabra model of turbulence with $\lambda = \sqrt{2+\sqrt{5}} \approx 2.058$. 
The continuous models allow understanding various properties of shell model solutions and provide their interpretation in physical space. We show 
that the asymptotic solutions of the dyadic model with Kolmogorov scaling correspond to the shocks (discontinuities) for the induced 
continuous solutions in physical space, and the finite-time blowup together with its 
viscous regularization follow the scenario similar to the Burgers equation. 
For the Sabra model, we provide the physical space representation for 
blowup solutions and intermittent turbulent dynamics.
\end{abstract}


\section{Introduction}

Shell models represent simplified ``toy'' models demonstrating various nontrivial phenomena of developed hydrodynamic turbulence, many of which are still not fully understood~\cite{frisch1995turbulence,biferale2003shell}. These models consider a geometric series of wavenumbers, 
$k_n  = k_0\lambda^n$ with $n \in \mathbb{Z}$ and fixed $\lambda > 1$. Each shell is 
represented by one or several (real or complex) numbers called shell speeds, which mimic 
the velocity field at a given scale $\delta r \sim 2\pi/k_n$. The evolution of shell speeds is governed by infinite-dimensional systems of 
ordinary differential equations, which allow for local interaction among the shells 
and must share several properties like scaling invariance, quadratic nonlinearity, 
energy conservation etc. with the Navier-Stokes equations or other hydrodynamic models. 
Originating in early 70s, see e.g. \cite{lorenz1972low,gledzer1973system,desnyansky1974evolution}, 
shell models became especially popular with the construction of the 
Gledzer--Ohkitani--Yamada (GOY) model~\cite{gledzer1973system,ohkitani1989temporal}, 
which has the chaotic intermittent dynamics in the inertial interval with the statistical 
properties close to the Navier-Stokes developed turbulence. 
In this paper, we consider the following shell models: the dyadic 
(Desnyansky--Novikov) model~\cite{desnyansky1974evolution} and the Sabra model 
of turbulence~\cite{l1998improved} representing a modified version of the GOY model.

The derivation of shell models is usually based of restricting the Fourier 
transformed hydrodynamic equations to a very limited number of modes (shells). 
Such simplification leads to a strongly 
reduced system, so that the same shell model can be derived from different 
original systems, e.g., Burgers or Navier-Stockes 
equations~\cite{waleffe2006some,mailybaev2012}. As a result, shell model 
solutions generally lose their quantitative relation with the original systems, 
though they may retain some important qualitative properties. This leads to the problem of 
interpreting the results obtained for shell models in their relation with the 
continuous solutions in physical space.

In this paper, we construct one-dimensional continuous hydrodynamic models, 
from which the shell models can be derived without any simplifications, i.e., 
in a rigorous way. This means that the Fourier transformed continuous models 
split into an infinite set of uncoupled subsystem, where each subsystem is 
equivalent to the same shell model under consideration. Such continuous models 
are characterized by nonlocal quadratic nonlinearity (similarly to the nonlocal 
term induced by pressure in incompressible flows), conserve energy and may 
have some other properties like the Hamiltonian structure etc. This construction 
is carried out for the two cases: the dyadic model with the intershell 
ratio $\lambda = 2^{3/2}$ and the Sabra model with $\lambda = \sqrt{2+\sqrt{5}} \approx 2.058$. 
Note that our approach uses the fixed values of $\lambda$, as opposed to the limiting models 
obtained as $\lambda \to 1$~\cite{andersen2000pulses}.

The proposed continuous models resolve the problem of representing the shell model 
solutions in physical space. This allows interpreting various results for shell 
models by relating them with the known properties of continuous flows or, more generally, 
of evolutionary partial differential equations. We demonstrate this by showing that 
the asymptotic solution with the Kolmogorov scaling, $u_n \propto k_n^{-1/3}$, 
in the dyadic shell model corresponds to a shock (discontinuity) for the induced continuous solution in 
physical space. Furthermore, the finite-time blowup together with its viscous 
regularization in the dyadic model follow the scenario similar to the 
Burgers equation for the continuous representation. Another implication of this approach is the physical space 
representation of intermittent turbulent dynamics in the Sabra model. 

The paper is organized as follows. Section~\ref{sec:Model} describes the properties of one-dimensional hydrodynamic models. Section~\ref{sec:DN} analyzes the continuous representation of the dyadic (Des\-nyan\-sky--Novikov) shell model, and Section~\ref{secSabra} does the same for the Sabra model of turbulence. We finish with some conclusions.

\section{One-dimensional hydrodynamic models}
\label{sec:Model}

We consider one-dimensional models for a scalar variable $u(x,t)$ in the form
\begin{equation}
\frac{\partial u}{\partial t}
+\frac{\partial g}{\partial x} = 
\nu
\frac{\partial^2u}{\partial x^2}+f,\quad 
x,t \in \mathbb{R},
\label{eq1}
\end{equation} 
where $\nu$ is a viscous coefficient, $f(x,t)$ is the forcing term and 
\begin{equation}
g(x,t) = \frac{1}{2\pi}\iint K(y-x,z-x)u(y,t)u(z,t)dydz
\label{eq2}
\end{equation} 
is the nonlocal quadratic flux term. For hydrodynamic models, where the quadratic term originates from the convective acceleration (and pressure for inviscid flows), it is natural to assume that $K(y,z)$ is a real homogeneous function of degree $-2$. Therefore, it can be considered in the form 
\begin{equation}
K(y,z) = \iint \varphi\left(\frac{p}{p+q}\right)e^{-i(py+qz)}dpdq,
\label{eq3}
\end{equation} 
with a real function $\varphi(\xi)$. For example, the product of Dirac 
delta functions $K(y,z) = \pi\delta(y)\delta(z)$ corresponds to $\varphi \equiv (4\pi)^{-1}$ and 
generates the Burgers equation with $g = u^2/2$ in Eq.~(\ref{eq1}). 

We do not specify the functional spaces 
for solutions $u(x,t)$ and for the kernel $K(y,z)$, assuming that they allow the 
integral (Fourier) transformations used below. We will comment on this issue when considering a specific form of $K(y,z)$ in the next section.
It is clear that the function $K(y,z)$ in Eq.~(\ref{eq2}) can always be chosen symmetric, i.e., $K(y,z) = K(z,y)$. One can check that permuting the variables $y \leftrightarrow z$ in the expression (\ref{eq3}) is equivalent to permuting $p \leftrightarrow q$ and substituting $\varphi(\xi)$ by $\varphi(1-\xi)$. Thus, the symmetry of $K(y,z)$ is equivalent to the condition
\begin{equation}
\varphi(\xi) = \varphi(1-\xi),
\label{eq3b}
\end{equation}
which will be assumed from now on. 
 
For the Fourier transformed function $u(k) = \int u(x)e^{-ikx}dx$, Eqs.~(\ref{eq1})--(\ref{eq3}) reduce to 
\begin{equation}
\frac{\partial u(k)}{\partial t}
= -ik\int \varphi\left(\frac{p}{k}\right)u(p)u(k-p)dp 
-\nu |k|^{2\alpha} u(k)+f(k),
\label{eq4}
\end{equation} 
where we omitted the argument $t$ for simplicity of notations. 
We also introduced the parameter $\alpha$, such that $\alpha = 1$ corresponds to 
Eq.~(\ref{eq1}) and $\alpha > 1$ determines the model with hyperviscosity. 
The mean value $\int u(x)dx$ is conserved by Eq.~(\ref{eq1}) provided that $\int f(x)dx = 0$ and $g \to 0$, $\partial u/\partial x \to 0$ as $|x| \to \infty$. 
We will assume the vanishing mean values, leading to $f(k) = u(k) = 0$ for $k = 0$. 
Recall the reality condition $u(-k) = u^*(k)$ for the Fourier transformed real function, 
where the asterisk denotes the complex conjugation.

\subsection{Energy conservation}

We define the energy as
\begin{equation}
E = \frac{1}{2}\int u^2(x)dx = \frac{1}{4\pi}\int |u(k)|^2dk.
\label{eq5}
\end{equation}
Let us show that the energy conservation condition in the inviscid model with zero force ($\nu = f = 0$) is given by the equality
\begin{equation}
\varphi\left(\xi\right) 
-\xi
\varphi\left(\frac{1}{\xi}\right) 
+(\xi-1)
\varphi\left(\frac{1}{1-\xi}\right) 
= 0
\label{eq6}
\end{equation} 
for all $\xi \in \mathbb{R}$. Indeed, using the reality condition $u^*(k) = u(-k)$ and $\nu = f = 0$ in Eqs.~(\ref{eq4}), (\ref{eq5}), we have
\begin{equation}
\frac{dE}{dt}
= \frac{1}{2\pi}
\mathrm{Re}\int u^*(k)\frac{\partial u(k)}{\partial t} dk
= \frac{1}{2\pi}\mathrm{Im}\iint k\varphi\left(\frac{p}{k}\right)u(-k)u(p)u(k-p)dpdk.
\label{eq8}
\end{equation}
By introducing $q = k-p$ and changing sign of $k$, we can write 
this expression as 
\begin{equation}
\frac{\partial E}{\partial t}
= -\frac{1}{2\pi}
\mathrm{Im}\iiint k\varphi\left(-\frac{p}{k}\right)\delta(k+p+q)u(k)u(p)u(q)dkdpdq.
\label{eq9}
\end{equation}

Due to the symmetry of the factor $u(k)u(p)u(q)$ with respect to permutations of variables $(k,p,q)$, 
this expression vanishes for an arbitrary function $u(k)$ if 
\begin{equation}
k\left[\varphi\left(-\frac{p}{k}\right) 
+\varphi\left(-\frac{q}{k}\right)\right] 
+p\left[\varphi\left(-\frac{k}{p}\right) 
+\varphi\left(-\frac{q}{p}\right)\right] 
+q\left[\varphi\left(-\frac{k}{q}\right) 
+\varphi\left(-\frac{p}{q}\right)\right] 
= 0
\label{eq10}
\end{equation} 
for $k+p+q = 0$. Dividing by $k$ and denoting 
\begin{equation}
\frac{p}{k} = -\xi,\quad  
\frac{q}{k} = \frac{-p-k}{k} = \xi-1,\quad  
\frac{p}{q} = \frac{\xi}{1-\xi},  
\label{eq11}
\end{equation} 
we write (\ref{eq10}) as
\begin{equation}
\varphi\left(\xi\right) 
+\varphi\left(1-\xi\right) 
-\xi\left[
\varphi\left(\frac{1}{\xi}\right) 
+\varphi\left(1-\frac{1}{\xi}\right)
\right] 
+(\xi-1)\left[
\varphi\left(\frac{1}{1-\xi}\right) 
+\varphi\left(\frac{\xi}{\xi-1}\right)
\right] 
= 0.
\label{eq6b}
\end{equation} 
This equation reduces to the form (\ref{eq6}) using the condition (\ref{eq3b}).

In the presence of force, similar derivations yield the equation
\begin{equation}
\frac{dE}{dt} = \int f(x)u(x)dx,
\label{eqE}
\end{equation} 
where the right-hand side can be interpreted as a work done by the force $f(x)$.

Note that the above derivation of energy conservation implies convergence of all the integrals. Otherwise, the system may dissipate the energy even if the condition (\ref{eq6}) is satisfied. For example, the condition (\ref{eq6}) is satisfied for the Burgers equations with $\varphi \equiv (4\pi)^{-1}$, while the shock (weak discontinuous) solutions are dissipative. In this case $u(k) \sim k^{-1}$ for large $k$ and the integral in Eq.~(\ref{eq8}) diverges.

\subsection{Hamiltonian structure}
Let us show that the condition 
\begin{equation}
\varphi\left(\frac{1}{\xi}\right)
+\varphi\left(\frac{1}{1-\xi}\right) = 2\varphi\left(\xi\right)
\label{eq12}
\end{equation} 
allows writing Eqs.~(\ref{eq1})--(\ref{eq3}) with $\nu = f = 0$ in the Hamiltonian form
\begin{equation}
\frac{\partial u}{\partial t} = \{u,H\},
\label{eq13}
\end{equation} 
where the Hamiltonian and the Poisson bracket are
\begin{equation}
H = \frac{1}{6\pi}\iiint K(y-x,z-x)u(x)u(y)u(z)dxdydz, 
\label{eq13c}
\end{equation} 
\begin{equation}
\{F,G\} = -\int \frac{\delta F}{\delta u}\frac{\partial}{\partial x} 
\frac{\delta G}{\delta u}dx.
\label{eq14}
\end{equation} 
Here we used the Poisson bracket known for the Korteweg--de Vries (KdV) and inviscid Burgers equations~\cite{gardner1971korteweg,Morrison1998}. Note that this Poisson bracket is noncanonical and has the Casimir invariant $C = \int udx$, which we set earlier to zero. 

In order to derive Eq.~(\ref{eq13}), we write expression (\ref{eq3}) in the form
\begin{equation}
K(y-x,z-x) = \iiint \varphi\left(-\frac{p}{k}\right)e^{-i(kx+py+qz)}
\delta(k+p+q) dpdqdk,
\label{eq3c}
\end{equation}
while similar expressions for $K(z-y,x-y)$ and $K(x-z,y-z)$ are obtained after substituting $\varphi\left(-\frac{p}{k}\right)$ by $\varphi\left(-\frac{q}{p}\right)$ and $\varphi\left(-\frac{k}{q}\right)$, respectively. Using these expressions with 
notations (\ref{eq11}) and condition (\ref{eq3b}), one can see that Eq.~(\ref{eq12}) implies 
\begin{equation}
K(z-y,x-y)+K(x-z,y-z) = 2K(y-x,z-x).
\label{eq16}
\end{equation} 
Then, for $F = u(x')$ and $G = H$ from Eq.~(\ref{eq13c}), we have 
\begin{equation}
\begin{array}{rcl}
\displaystyle
\frac{\delta F}{\delta u(x)} & = & \delta(x-x'),
\\[12pt]
\displaystyle
\frac{\delta G}{\delta u(x)} & = &  
\displaystyle
\frac{1}{6\pi}\iint 
\left[K(y-x,z-x)+K(z-y,x-y)+K(x-z,y-z)\right]
u(y)u(z)dydz
\\[15pt]
& = &  
\displaystyle
\frac{1}{2\pi}\iint 
K(y-x,z-x)
u(y)u(z)dydz,
\end{array}
\label{eq15}
\end{equation} 
where the last equality follows from Eq.~(\ref{eq16}).
Substituting these expressions into Eqs.~(\ref{eq13}), (\ref{eq14}) yields Eqs.~(\ref{eq1}), (\ref{eq2}) with $\nu = f = 0$. 

\section{Continuous representation of the Desnyansky--Novikov model}
\label{sec:DN}

Let us consider the function
\begin{equation}
\varphi(\xi) 
= \frac{1}{2}\delta\left(\xi-\frac{1}{2}\right)
+2\delta(\xi+1)+2\delta(\xi-2),
\label{eqD1}
\end{equation}
which is the sum of three Dirac delta functions. 
For the physical space representation of model (\ref{eq1}), (\ref{eq2}), we should find the kernel $K(y,z)$ given by Eq.~(\ref{eq3}) with the function (\ref{eqD1}). The integrals in Eq.~(\ref{eq3}) can be taken explicitly using the relation 
\begin{equation}
\delta\left(\frac{\xi-\xi_0}{a(\xi)}\right) = |a(\xi_0)|\delta(\xi-\xi_0). 
\label{eqD1e}
\end{equation}
For the first term, we obtain
\begin{equation}
\begin{array}{c}
\displaystyle
\int \frac{1}{2}\delta\left(\frac{p}{p+q}-\frac{1}{2}\right)
e^{-i(py+qz)}dpdq
= \int \frac{1}{2}\delta\left(\frac{p-q}{2(p+q)}\right)
e^{-i(py+qz)}dpdq
\\[15pt]
\displaystyle
= 2\int |q|e^{-iq(y+z)}dq
= -\frac{4}{(y+z)^2},
\end{array}
\label{eqD1d}
\end{equation}
where we used that the Fourier transform of $|x|$ is the generalized function $-2k^{-2} = 2\frac{d}{dk}\left[\mathrm{p.v.\left(\frac{1}{k}\right)}\right]$, see e.g.~\cite{kanwal2004generalized}.
Integrating similarly the other terms in Eqs.~(\ref{eq3}), (\ref{eqD1}), yields 
\begin{equation}
K(y,z) 
= -\frac{4}{(y+z)^2}
-\frac{4}{(y-2z)^2}
-\frac{4}{(z-2y)^2},
\label{eqD1c}
\end{equation}
and one should consider the singular integrals in Eq.~(\ref{eq2}) with the Hadamard regularization.

The Fourier transformed continuous model has a simpler form. Indeed, substituting $\varphi(\xi)$ from Eq.~(\ref{eqD1}) into Eq.~(\ref{eq4}) and using Eq.~(\ref{eqD1e}) yields
\begin{equation}
\frac{\partial u(k)}{\partial t}
 = -ik|k|\left[
\frac{1}{2}u^2\left(\frac{k}{2}\right)
+4u^*(k)u(2k)
\right]
-\nu |k|^{2\alpha}u(k)+f(k),
\label{eqD2}
\end{equation} 
where we used the reality condition $u(-k) = u^*(k)$. We will show now that this equation is equivalent to an infinite set of uncoupled discrete (shell) models.

Let us define the geometric progression
\begin{equation}
k_n = k_0\lambda^{n},\quad
\lambda = 2^{3/2},\quad
n \in \mathbb{Z},
\label{eqD3}
\end{equation} 
and the corresponding variables
\begin{equation}
u_n = -ik_n^{1/3}u\left(k_n^{2/3}\right),\quad
f_n = -ik_n^{1/3}f\left(k_n^{2/3}\right). 
\label{eqD4}
\end{equation} 
Then Eq.~(\ref{eqD2}) taken for $k = k_n^{2/3}$ reduces to the form
\begin{equation}
\frac{\partial u_n}{\partial t}
= k_n u_{n-1}^2
-k_{n+1}u_n^*u_{n+1}
-\nu_nu_n+f_n,
\quad n \in \mathbb{Z},
\label{eqD5}
\end{equation} 
where introduced the viscous factors 
\begin{equation}
\nu_n = \nu k_n^{4\alpha/3}.
\label{eqD6}
\end{equation} 
Equation (\ref{eqD5}) is the shell model, where $k_n$ is the shell wavenumber (forming a geometric progression in $n$), $u_n \in \mathbb{C}$ is the complex  shell speed, and the interaction occurs between the neighboring shells.

One may consider the real variables $u_n \in \mathbb{R}$. According to relation (\ref{eqD4}), 
this is the case when 
$u(k)$ is a purely imaginary function and, 
hence, the solution of the continuous model is an odd function in physical space, $u(-x) = -u(x)$.
For real variables, system (\ref{eqD5}) becomes
\begin{equation}
\frac{\partial u_n}{\partial t}
= k_n u_{n-1}^2
-k_{n+1}u_nu_{n+1}
-\nu_nu_n+f_n,
\quad n \in \mathbb{Z}.
\label{eqD5b}
\end{equation} 
This system is known as the Desnyansky--Novikov (DN) 
shell model~\cite{desnyansky1974evolution}, also called the dyadic shell model. 
Note that, due to the exponent $4\alpha/3$ in this viscous term (\ref{eqD6}), the conventional choice of $\nu_n = \nu k_n^2$ in the shell model 
corresponds to the continuous model (\ref{eqD2}) with hyperviscosity given by $\alpha = 3/2$. 

We have shown that the hydrodynamic model given by Eqs.~(\ref{eq1})--(\ref{eq3})
and (\ref{eqD1}) splits into a set of equivalent infinite-dimensional subsystems (\ref{eqD5}). 
Each of these subsystems corresponds to a specific value of the parameter $k_0$ in Eq.~(\ref{eqD3}), which must be taken in the interval 
\begin{equation}
1 \le k_0 < \lambda.
\label{eqD8}
\end{equation} 
For odd continuous solutions, $u(-x) = -u(x)$, these subsystems take the form of 
the DN shell model (\ref{eqD5b}), 
where the real variables $u_n$ are related to $u(k)$ by Eq.~(\ref{eqD4}). 

\subsection{Energy conservation and Hamiltonian form}

Let us consider the shell model (\ref{eqD5}) in the inviscid unforced case, 
i.e., when $\nu_n = f_n = 0$ for all $n$. Using Eq.~(\ref{eqD1e}), 
one can check that the function (\ref{eqD1}) satisfies the conditions 
\begin{equation}
\varphi\left(\xi\right)
= \varphi\left(1-\xi\right)
= \varphi\left(\frac{1}{\xi}\right)
= \varphi\left(\frac{1}{1-\xi}\right),
\label{eq17}
\end{equation} 
which  imply the simultaneous energy conservation and Hamiltonian structure, 
see Eqs.~(\ref{eq3b}), (\ref{eq6}) and (\ref{eq12}).

Equations (\ref{eqD3}) and (\ref{eqD4}) with $k = k_n^{2/3}$ yield
\begin{equation}
u(k) = ik_n^{-1/3}u_n(k_0),\quad 
k = k_n^{2/3} = k_0^{2/3}\lambda^{2n/3},
\label{eqD8e}
\end{equation} 
\begin{equation}
dk = dk_n^{2/3} = \frac{2}{3}\lambda^{2n/3}k_0^{-1/3}dk_0 
= \frac{2}{3}k_n^{2/3}k_0^{-1}dk_0,
\label{eqD8b}
\end{equation}
where we specified the parameter $k_0$ explicitly as an argument for the shell speed $u_n$. 
Using these expressions with the interval (\ref{eqD8}) for $k_0$, 
the energy (\ref{eq5}) can be written as
\begin{equation}
E = \frac{1}{4\pi}\int \left|u(k)\right|^2dk 
= \frac{1}{4\pi}\int_0^\lambda  
\frac{2}{3}k_0^{-1} \sum_{n \in \mathbb{Z}} \left|u_n(k_0)\right|^2dk_0
= \frac{1}{6\pi}\int_0^\lambda  
k_0^{-1} E_s(k_0) dk_0,
\label{eqD8c}
\end{equation} 
where 
\begin{equation}
E_{s}(k_0) = \sum_{n \in \mathbb{Z}} \left|u_n(k_0)\right|^2.
\label{eqD8d}
\end{equation} 
These expressions relate the energy $E$ of the continuous model with the energies $E_s$ of the shell models for all $k_0$ from the interval (\ref{eqD8}). 

Note that the formal condition of energy conservation was used for derivation 
of the DN shell model~\cite{desnyansky1974evolution}; 
see also \cite{cheskidov2007inviscid,cheskidov2008blow,cheskidov2010inviscid} 
for the detailed analysis in the case of $\lambda = 2^{5/2}$.
If the initial condition at $t = 0$ 
has finite norm $\sum_{n \in \mathbb{Z}} k_n^2|u_n|^2 < \infty$, 
then the inviscid shell model has a unique solution for small times 
and the energy $E_s$ is conserved, see e.g. \cite{constantin2007regularity,mailybaev2012c}. 
However, this norm blows up in finite time 
leading to dissipative solutions, which we discuss in the next section. 
 
The inviscid unforced continuous model has the Hamiltonian structure given 
by Eqs.~(\ref{eq13})--(\ref{eq14}). This structure induces the corresponding 
Hamiltonian form for shell model (\ref{eqD5}). 
The direct derivation of this Hamiltonian form from the continuous formulation 
is lengthy, and it is easy to find it directly taking into account that the 
Hamiltonian must be a cubic function of shell speeds $u_n$. Indeed, 
consider the shell model equations written in the canonical form
\begin{equation}
\frac{\partial a_n}{\partial t} = -i\frac{\partial H}{\partial a_n^*}
= \frac{1}{2}k_n^{4/3}a_{n-1}^2-k_{n+1}^{4/3}a_n^*a_{n+1},
\label{eqD10}
\end{equation} 
where $\{a_n,a_n^*\}$, $n \in \mathbb{Z}$, are pairs of 
complex canonical variables and the real Hamiltonian is given by
\begin{equation}
H = \frac{i}{2}\sum_n k_n^{4/3}\left(a_{n-1}^2a_n^*-(a^*_{n-1})^2a_n\right). 
\label{eqD11}
\end{equation} 
It is easy to see that Eq.~(\ref{eqD10}) reduces to Eq.~(\ref{eqD5}) 
with $\nu_n = f_n = 0$ for $a_n = k_n^{-1/3}u_n$, 
where one should use the relation $k_{n+1}^{2/3} = 2k_n^{2/3}$ following from Eq.~(\ref{eqD3}).

\subsection{Shock wave solutions in the inviscid model}

Let us consider real solutions of the inviscid DN model (\ref{eqD5b}) with the specific constant 
forcing applied to the shell with $n = 0$, i.e.,
\begin{equation}
\frac{\partial u_n}{\partial t}
= k_n u_{n-1}^2
-k_{n+1}u_nu_{n+1}+f_n,\quad f_n = 2k_0^{1/3}\delta_{n0}.
\label{eqD5i}
\end{equation} 
Here $\delta_{n0}$ is the Kronecker delta, which is equal to $1$ for $n = 0$ and zero otherwise.
We will consider initial conditions in the form
\begin{equation}
t = 0:\quad u_n = u_n^0\quad \textrm{for}\quad n \ge 0; \quad
u_n = 0\quad \textrm{for}\quad n < 0,
\label{eqD15}
\end{equation} 
assuming no perturbation of the shells with $n < 0$ corresponding to large scales. 
In this case, $u_n \equiv 0$ for $n < 0$ and $t \ge 0$. 
Thus, the dynamics is restricted to the shells with $n \ge 0$.

\begin{figure}
\centering
\includegraphics[width = 0.99\textwidth]{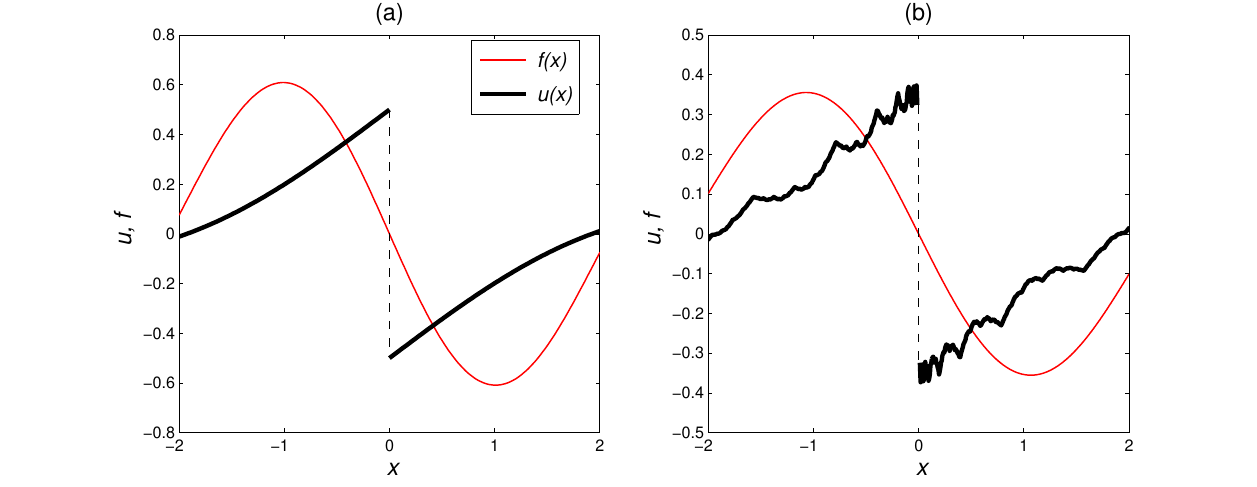}
\caption{Stationary shock wave solution $u(x)$ (bold line) and forcing $f(x)$ (thin line) for the inviscid continuous model. This solution corresponds to the DN shell model for: (a) forcing $f_n = 2k_0^{1/3}\delta_{n0}$ with the stationary solution (\ref{eqD16}); (b) forcing $f_n = 2k_0^{-1}\delta_{n0}$ with the stationary solution (\ref{eqF2}).}
\label{fig1}
\end{figure}

One can check by direct substitution that 
\begin{equation}
u_n = k_n^{-1/3},\quad n \ge 0,
\label{eqD16}
\end{equation}
is a fixed-point solution of Eq.~(\ref{eqD5i}), where $k_n = k_0\lambda^n$ 
with $\lambda = 2^{3/2}$. This family of stationary solutions $\{u_n\}_{n\in\mathbb{Z}}$ 
defined for each $k_0$ from the interval (\ref{eqD8})
induces a stationary solution $u(x)$ of the corresponding continuous model (\ref{eq1}), 
(\ref{eq2}) and (\ref{eqD1c}) with vanishing 
viscosity $\nu = 0$ and specific constant forcing 
$f(x)$. The Fourier transformed continuous solution $u(k)$ and 
forcing $f(k)$ can be found using 
Eqs.~(\ref{eqD3}) and (\ref{eqD4}) with $k = k_n^{2/3}$ as 
\begin{equation}
u(k) = ik^{-1/2}u_n,\quad
f(k) = ik^{-1/2}f_n,\quad
k = k_n^{2/3} = k_0^{2/3} 2^n.
\label{eqD14}
\end{equation} 
Recall that $u(k) = 0$ was assumed for $k = 0$. Also, $u_n = 0$ for negative $n$, 
i.e., $u(k) = 0$ for $0 \le k < 1$. Using 
$f_n$ from Eq.~(\ref{eqD5i}) and the solution (\ref{eqD16}) in Eq.~(\ref{eqD14}), we obtain
\begin{equation}
u(k) = \left\{\begin{array}{ll}                      
ik^{-1}, & |k| \ge 1;\\[5pt]
0, & |k| < 1;
\end{array}\right.
\quad 
f(k) = \left\{\begin{array}{ll}                      
2i\,\mathrm{sgn}(k), & 1\le |k| < 2;\\[5pt]
0, & |k| \notin [1,\,2);
\end{array}\right.
\label{eqD17}
\end{equation}
where the values for negative $k$ are obtained from the reality condition 
$u(-k) = u^*(k)$. 
The physical space solution $u(x)$ and forcing $f(x)$ are 
found using the inverse Fourier transform $u(x) = \frac{1}{2\pi}\int u(k)e^{ikx}dk$ as
\begin{equation}
u(x) = \frac{\mathrm{si}|x|}{\pi}\,\mathrm{sgn}\,x,
\quad 
f(x) = -\frac{4}{\pi x}\sin \frac{3x}{2}\sin \frac{x}{2},
\label{eqD18}
\end{equation}
where $\mathrm{si}(x) = -\int_x^\infty \frac{\sin t}{t}dt$ is the sine integral function. 
The functions (\ref{eqD18}) are shown in Fig.~\ref{fig1}(a). The forcing $f(x)$ 
is analytic, while the solution $u(x)$ has a discontinuity 
at $x = 0$. Thus, the stationary shell model solution (\ref{eqD16}) corresponds 
to the standing shock wave solution (\ref{eqD18}) for the continuous model.

It is instructive to consider a different forcing model, e.g.,
\begin{equation}
f_n = 2k_0^{-1}\delta_{n0}.
\label{eqF1}
\end{equation}
The fixed-point solution of Eq.~(\ref{eqD5i}) with the new forcing term (\ref{eqF1}) is
\begin{equation}
u_n = k_0^{-2/3}k_n^{-1/3},\quad n \ge 0.
\label{eqF2}
\end{equation}
The Fourier transformed forcing $f(k)$ and solution $u(k)$ are given by Eq.~(\ref{eqD14}). 
The corresponding functions $f(x)$ and $u(x)$ in physical space can 
found analytically or numerically, e.g., 
using the inverse Fourier transform method described in~\cite{bailey1994fast}. 
These functions are shown in Fig.~\ref{fig1}(b). 
The forcing $f(x)$ is an analytic function, while the solution $u(x)$ has a discontinuity at the origin. For $x \ne 0$, the function $u(x)$ is continuous with the fractal-like shape.

The shell model solution (\ref{eqD16}) has the infinite norm 
$\sum_{n\in\mathbb{Z}}k_n^2u_n^2 = \infty$ and it
is dissipative~\cite{cheskidov2007inviscid}, 
i.e., it does not satisfy 
the energy conservation relation (\ref{eqE}) even though the 
model has no viscosity, $\nu = 0$. Also, the solution (\ref{eqD16}) 
was shown (in the case of $\lambda = 2^{5/2}$) to be a global attractor for  
finite energy initial conditions~\cite{cheskidov2007inviscid}. 
These two facts get the clear interpretation in physical space: 
the global attractor is a shock wave, which represents the dissipative mechanism. 
This mechanism is similar to that known for weak solutions 
(shocks) of inviscid scalar conservation 
laws with the flux function $g = g(u)$ in Eq.~(\ref{eq1}), e.g., 
for the inviscid Burgers equation.  
 
As we mentioned earlier, solutions of the inviscid DN model are unique and conserve the energy 
(in the absence of forcing), if the norm $\sum_{n\in\mathbb{Z}} k_n^2u_n^2 < \infty$.
This norm becomes infinite (blows up) in finite time~\cite{dombre1998intermittency,katz2005finite,cheskidov2007inviscid}. 
The asymptotic blowup structure is self-similar and universal, and 
it can be described using the renormalization 
technique~\cite{dombre1998intermittency,mailybaev2012c}. 
Following the derivations similar to Eqs.~(\ref{eqD8e})--(\ref{eqD8d}), 
one can relate the above norm with the integral $\int |k|^3|u(k)|^2dk$ 
for the Fourier transformed continuous solution. 
This suggests that the physical space solution $u(x,t)$ should be understood 
in the strong sense before the blowup and in the weak sense after the blowup similarly to 
the shell models~\cite{constantin2007regularity}; 
these questions require further elaboration and are beyond the scope of this paper. 

\subsection{Viscous solutions}
\label{sec:DNvisc}

Now let us consider the viscous DN shell model (\ref{eqD5b}). We assume 
the stationary forcing terms (\ref{eqF1}) and choose the viscous coefficients in the form
\begin{equation}
\nu_n = \nu k_0^{-4/3}k_n^{4/3} = \nu \lambda^{4n/3},
\label{eqD6b}
\end{equation} 
where $\nu > 0$ is a small viscosity parameter. The viscous coefficients (\ref{eqD6b}), whose original form was given in Eq.~(\ref{eqD6}), are modified in order to facilitate the construction of the 
continuous solution $u(x,t)$. One can expect that such changes are only 
important at viscous scales (large $k$), leading to the same behavior in the vanishing 
viscosity limit $\nu \to +0$ \cite{lvov_universal1998}. 

Substituting Eqs.~(\ref{eqF1}) and (\ref{eqD6b}) into the DN shell model (\ref{eqD5b}) 
with $k_n = k_0\lambda^n$, yields
\begin{equation}
\frac{\partial U_n}{\partial t}
= \lambda^n U_{n-1}^2
-\lambda^{n+1}U_nU_{n+1}
-\nu \lambda^{4n/3}U_n
+2\delta_{n0},
\quad 
U_n = k_0u_n,
\label{eqD5c}
\end{equation}
where we multiplied both sides by $k_0$.
This representation allows constructing particular solutions 
$\{u_n(k_0,t)\}_{n \ge 0}$ for any value of the parameter $k_0$ 
using a single solution $\{U_n(t)\}_{n \ge 0}$. 
As the initial condition at $t = 0$, we will consider vanishing shell velocities 
$U_n(0) = 0$ for all shells $n$. 

\begin{figure}
\centering
\includegraphics[width = 0.9\textwidth]{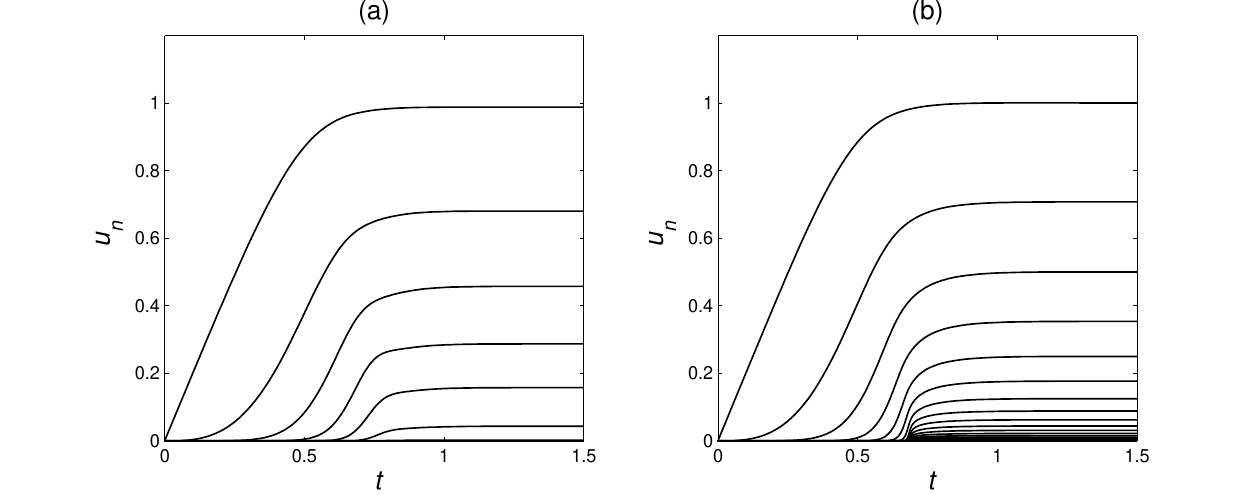}
\caption{Solutions of the DN shell model with constant forcing (\ref{eqF1}) 
and zero initial conditions: (a) viscous coefficient $\nu = 0.1$, 
(b) viscous coefficient $\nu = 10^{-5}$. }
\label{fig2}
\end{figure}

Due to the viscous coefficients (\ref{eqD6b}), which grow as $\lambda^{4n/3} = 2^{2n}$, 
the shell speeds decay rapidly for large $n$. 
Thus, the numerical solution can be obtained with high accuracy using a 
finite number of shell velocities $u_n$ with $n = 0,\ldots,N$ 
and assuming vanishing velocities for other shells (for example, for $\nu = 10^{-4}$, 
one has $u_{20} \sim 10^{-25}$). 
The numerical solutions are shown in Fig.~\ref{fig2}. 
One can observe the special behavior near $t_* \approx 0.691$ in Fig.~\ref{fig2}(a), 
which corresponds to the viscous coefficient $\nu = 0.1$. 
In the vanishing viscosity limit $\nu \to +0$, the point $t_*$ corresponds 
to a finite time blowup, see Fig.~\ref{fig2}(b). For larger times, 
a steady state is formed. 

\begin{figure}
\centering
\includegraphics[width = 0.99\textwidth]{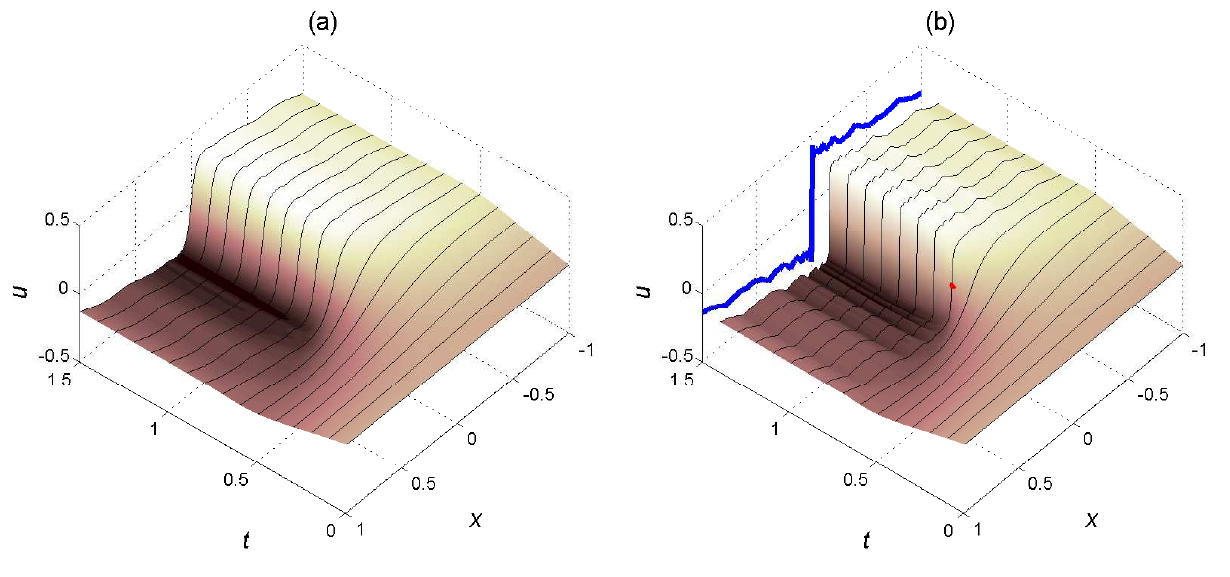}
\caption{Physical space solutions $u(x,t)$ induced by 
solutions of the viscous DN shell model with constant forcing (\ref{eqF1}) 
and zero initial conditions: (a) viscous coefficient $\nu = 0.1$, 
(b) viscous coefficient $\nu = 0.01$. The red point in the right figure shows 
the location of blowup for the inviscid system. Also, the right 
figure shows the stationary solution of the inviscid model 
(bold blue line located at $t = 1.5$), see Fig.~\ref{fig1}(b).}
\label{fig3}
\end{figure}

For constructing the corresponding solution of the continuous model, 
we use Eqs.~(\ref{eqD8e}) and (\ref{eqD5c}), 
which yield the Fourier transformed solution $u(k,t)$.
The inverse Fourier transform (performed numerically 
as described in~\cite{bailey1994fast}) yields the solution $u(x,t)$ in physical space.
Figure~\ref{fig3} shows the solution $u(x,t)$ corresponding to the DN model 
with two different viscous coefficients, $\nu = 0.1$ and $\nu = 0.01$. 
These solutions demonstrate the formation of a smooth standing wave. 
For small viscosity, this solution converges to the inviscid solution 
described earlier, i.e., to the 
finite time blowup followed by formation of a shock wave. The blowup point corresponds 
to $x = u = 0$ and $t_* \approx 0.691$, as obtained from the shell model solution 
in Fig.~\ref{fig2}(b). This point is shown in Fig.~\ref{fig3}(b) by a bold red point. 
Also, for comparison, this figure shows the stationary solution of the inviscid model 
(bold blue line located at $t = 1.5$), which was obtained earlier in Fig.~\ref{fig1}(b). 

We conclude that physical space solutions $u(x,t)$, 
which are induced by solutions of the DN shell model, 
are similar to solutions of scalar conservation laws: a shock wave is formed in finite time in the inviscid model, while the viscosity transforms this shock into a smooth wave solution.

\section{Continuous representation of the Sabra model}
\label{secSabra}

In this section, we consider the one-dimensional 
hydrodynamic model (\ref{eq1})--(\ref{eq3}) given by the function  
\begin{equation}
\varphi(\xi) = \frac{\psi(\xi)+\psi(1-\xi)}{2},\quad
\psi(\xi) 
= -\sigma^{3}\delta(\xi-\sigma^{2})
+(1+c)\delta(\xi-\sigma)
+c\sigma^{-3}\delta\left(\xi-\sigma^{-1}\right),
\label{eqB1}
\end{equation}
where $c$ is a constant real parameter and $\sigma = (1+\sqrt{5})/2 \approx 1.618$ is 
the golden ratio satisfying the equation \begin{equation}
1+\sigma = \sigma^2.
\label{eqB0}
\end{equation} 
After substituting expression (\ref{eqB1}) into Eq.~(\ref{eq3}), the lengthy 
derivations similar to Eq.~(\ref{eqD1d}) with the 
use of Eq.~(\ref{eqB0}) yield the kernel of the continuous model in the form
\begin{equation}
K(y,z) 
= K_\psi(y,z)+K_\psi(z,y),\quad
K_\psi(y,z) 
= \frac{\sigma}{(\sigma y-z)^2}-\frac{(1+c)\sigma^2}{(\sigma^2y-z)^2}
-\frac{c\sigma}{(\sigma y+z)^2}.
\label{eqB1c}
\end{equation}
Singular integrals in Eq.~(\ref{eq2}) must be taken with the Hadamard regularization.

The Fourier transformed continuous model is obtained by substituting 
function (\ref{eqB1}) into Eq.~(\ref{eq4}). Using relations (\ref{eqD1e}) and (\ref{eqB0}), this yields
\begin{equation}
\begin{array}{rcl}
\displaystyle
\frac{\partial u(k)}{\partial t}
& = &\displaystyle
-ik|k|\left(
-\sigma^{3}u(\sigma^{2}k)u(-\sigma k)
+(1+c)u(\sigma k)u(-\sigma^{-1} k)
+c\sigma^{-3}u(\sigma^{-1} k)u(\sigma^{-2} k)
\right)\\[12pt]
&&\displaystyle
-\nu |k|^{2\alpha}u(k)+f(k).
\end{array}
\label{eqB2}
\end{equation} 

Now let us define the geometric progression
\begin{equation}
k_n = k_0\lambda^{n},\quad
\lambda = \sigma^{3/2} = \sqrt{2+\sqrt{5}} \approx 2.058, \quad
n \in \mathbb{Z},
\label{eqB3}
\end{equation} 
with $1 \le k_0 < \lambda$, and introduce the corresponding variables
\begin{equation}
u_n = k_n^{1/3}u\left(k_n^{2/3}\right),\quad
f_n = k_n^{1/3}f\left(k_n^{2/3}\right). 
\label{eqB4}
\end{equation} 
Then Eq.~(\ref{eqB2}) taken for $k = k_n^{2/3} = k_0^{2/3}\sigma^n$ reduces to the form
\begin{equation}
\frac{\partial u_n}{\partial t}
=
i\left[
k_{n+1}u_{n+2}u^*_{n+1}
-(1+c)k_nu_{n+1}u^*_{n-1}
-ck_{n-1}u_{n-1}u_{n-2}
\right]
-\nu_nu_n+f_n,
\quad n \in \mathbb{Z},
\label{eqB7}
\end{equation} 
where we used the reality condition $u(-k) = u^*(k)$ for negative arguments; 
the viscous coefficients $\nu_n = \nu k_n^{4\alpha/3}$ are the same 
as in Eq.~(\ref{eqD6}). The system (\ref{eqB7}) 
is known as the Sabra shell model of turbulence~\cite{l1998improved}. 
The most popular choice of the intershell ratio in the Sabra model is $\lambda = 2$, 
which is very close to the value $\lambda \approx 2.058$ defined by the continuous model. 

We have shown that the hydrodynamic model given by Eqs.~(\ref{eq1})--(\ref{eq3})
and (\ref{eqB1}) splits into infinite-dimensional subsystems (\ref{eqB7}). 
Each of these subsystems corresponds to a specific value of the parameter $k_0$ from the interval (\ref{eqD8}) and represents the Sabra shell model. Note that the conventional choice of the viscous term $\nu_n = \nu k_n^2$ in the Sabra model corresponds to the exponent $\alpha = 3/2$, i.e., to the continuous model with hyperviscosity. 

\subsection{Energy conservation and Hamiltonian form}

This section contains technical derivations verifying the conditions (\ref{eq6}) 
and (\ref{eq12}) for the energy conservation and Hamiltonian structure in the 
continuous model, leading to the analogous properties of the induced Sabra model.
 
Using Eqs.~(\ref{eqB1}), (\ref{eqB0}) and (\ref{eqD1e}), one can check that
\begin{eqnarray}
\psi\left(1-\xi\right)
& = & -\sigma^{3}\delta(\xi+\sigma)
+(1+c)\delta(\xi+\sigma^{-1})
+c\sigma^{-3}\delta(\xi-\sigma^{-2}),
\label{eqB1x}
\\[5pt]
\psi\left(\frac{1}{\xi}\right)
& = & -\sigma^{-1}\delta(\xi-\sigma^{-2})
+(1+c)\sigma^{-2}\delta(\xi-\sigma^{-1})
+c\sigma^{-1}\delta(\xi-\sigma),
\label{eqC1}
\\[5pt]
\psi\left(1-\frac{1}{\xi}\right)
& = & -\sigma\delta(\xi+\sigma^{-1})
+(1+c)\sigma^{2}\delta(\xi+\sigma)
+c\sigma\delta(\xi-\sigma^{2}),
\label{eqC2}
\\[5pt]
\psi\left(\frac{1}{1-\xi}\right)
& = & -\sigma^{-1}\delta(\xi-\sigma^{-1})
+(1+c)\sigma^{-2}\delta(\xi-\sigma^{-2})
+c\sigma^{-1}\delta(\xi+\sigma^{-1}),
\label{eqC3}
\\[5pt]
\psi\left(\frac{\xi}{\xi-1}\right)
& = & -\sigma\delta(\xi-\sigma)
+(1+c)\sigma^{2}\delta(\xi-\sigma^2)
+c\sigma\delta(\xi+\sigma),
\label{eqC4}
\\[5pt]
-\xi\psi\left(\frac{1}{\xi}\right)
& = & \sigma^{-3}\delta(\xi-\sigma^{-2})
-(1+c)\sigma^{-3}\delta(\xi-\sigma^{-1})
-c\delta(\xi-\sigma),
\label{eqB1b}
\\[5pt]
-\xi\psi\left(1-\frac{1}{\xi}\right)
& = & -\delta(\xi+\sigma^{-1})
+(1+c)\sigma^{3}\delta(\xi+\sigma)
-c\sigma^{3}\delta(\xi-\sigma^{2}),
\label{eqB1f}
\\[5pt]
(\xi-1)\psi\left(\frac{1}{1-\xi}\right)
& = & \sigma^{-3}\delta(\xi-\sigma^{-1})
-(1+c)\sigma^{-3}\delta(\xi-\sigma^{-2})
-c\delta(\xi+\sigma^{-1}),
\label{eqB1d}
\\[5pt]
(\xi-1)\psi\left(\frac{\xi}{\xi-1}\right)
& = & -\delta(\xi-\sigma)
+(1+c)\sigma^{3}\delta(\xi-\sigma^{2})
-c\sigma^{3}\delta(\xi+\sigma).
\label{eqB1g}
\end{eqnarray}
Since $\varphi(\xi) = \frac{\psi(\xi)+\psi(1-\xi)}{2}$, the energy conservation 
condition (\ref{eq6}) reduces to checking that the sum of expression (\ref{eqB1}) 
for $\psi(\xi)$ and Eqs.~(\ref{eqB1x}), (\ref{eqB1b})--(\ref{eqB1g}) gives zero. 
It is straightforward to check that this condition holds indeed. The same expressions 
(\ref{eqD8c}) and (\ref{eqD8d})
hold for the energy of the continuous system $E$ and 
for the energy of the Sabra model $E_s$.

Energy conservation was one of the criteria for the construction of the Sabra 
shell model~\cite{l1998improved}. The unforced inviscid Sabra 
model (\ref{eqB7}) possesses another quadratic invariant 
\begin{equation}
H_s = \sum_{n\in \mathbb{Z}} c^{-n}|u_n|^2. 
\label{eqH}
\end{equation} 
This invariant is associated with the helicity for $c < 0$ (not sign-definite invariant) 
and with the enstrophy for $c > 0$ (sign definite invariant); 
these definitions are physically relevant for $0 < |c| < 1$.

The Hamiltonian structure condition (\ref{eq12}) with the function (\ref{eqB1}) takes the form
\begin{equation}
\psi\left(\frac{1}{\xi}\right)
+\psi\left(1-\frac{1}{\xi}\right)
+\psi\left(\frac{1}{1-\xi}\right) 
+\psi\left(\frac{\xi}{\xi-1}\right) 
= 2\psi\left(\xi\right)+2\psi\left(1-\xi\right).
\label{eq12b}
\end{equation} 
Using Eqs.~(\ref{eqB1}) and (\ref{eqB1x})--(\ref{eqC4}), 
it is straightforward to check that the condition (\ref{eq12b}) is satisfied if 
and only if $c = -\sigma^2 \approx -2.618$. Indeed,  
using Eqs.~(\ref{eqB0}) and (\ref{eqB3}),
one can check that the Sabra model (\ref{eqB7}) with $\nu_n = f_n = 0$ and $c = -\sigma^2$ 
can be written 
in the canonical Hamiltonian form
\begin{equation}
\frac{\partial a_n}{\partial t} = -i\frac{\partial H}{\partial a_n^*}
= i k_n^{4/3}\left(\sigma^{2} a_{n+2}a_{n+1}^*+a_{n+1}a_{n-1}^*
+\sigma^{-2}a_{n-1}a_{n-2}\right),
\label{eqH1}
\end{equation} 
where $a_n = \sigma k_n^{-1/3}u_n$, and $\{a_n,a_n^*\}$ are pairs of 
complex canonical variables with the Hamiltonian 
\begin{equation}
H = -\sum_n k_n^{4/3}\left(a_{n-1}^*a_n^*a_{n+1}+a_{n-1}a_na_{n+1}^*\right). 
\label{eqH2}
\end{equation} 

Note that a different Hamiltonian representation of the Sabra model was found in~\cite{l1999hamiltonian} for $c = -\sigma^{-2} \approx -0.382$. It has the form
\begin{equation}
\frac{\partial a_n}{\partial t} = -i\frac{\partial H}{\partial a_n^*},\quad
a_n = \left\{\begin{array}{ll}
\sigma^{-1} k_n^{1/3}u_n,& \textrm{even }n;\\[3pt]
-\sigma^{-1} k_n^{1/3}u_n^*,& \textrm{odd }n;
\end{array}
\right.
\label{eqH3}
\end{equation} 
with the Hamiltonian 
\begin{equation}
H = \sum_n k_n^{2/3}\left(a_{n-1}a_n^*a_{n+1}^*+a_{n-1}^*a_na_{n+1}\right). 
\label{eqH4}
\end{equation}

\subsection{Inviscid model}

Let us consider the unforced inviscid Sabra model
\begin{equation}
\frac{\partial u_n}{\partial t}
=
i\left[
k_{n+1}u_{n+2}u^*_{n+1}
-(1+c)k_nu_{n+1}u^*_{n-1}
-ck_{n-1}u_{n-1}u_{n-2}
\right],\quad n \ge 0.
\label{eqS1}
\end{equation} 
Here we assume that all shell speeds $u_n$ with $n < 0$ vanish. 
This can be arranged, for example, by considering vanishing initial conditions for 
shells with $n < 0$ and specifying the forcing terms $f_{-2}$ and $f_{-1}$ such 
that the right-hand sides of the equations for $u_{-2}$ and $u_{-1}$ vanish.  

First let us consider the case $c = \lambda^{-2} \approx 0.236$, when  
the conserved quantity (\ref{eqH}) is positive and can be 
associated with the enstrophy. For this reason, 
Eq.~(\ref{eqS1}) with $c = \lambda^{-2}$ is considered to be the shell model 
for two-dimensional inviscid flow (2D Euler equations). This model has a unique 
global in time solution for initial conditions of finite energy and 
enstrophy~\cite{constantin2007regularity}. 

Let us consider the complex initial conditions of the form
\begin{equation}
t = 0:\quad u_0 = ik_0^{-1},\quad u_1 = 2k_0^{-1},\quad u_n = 0,\quad n \ge 2.
\label{eqS2}
\end{equation} 
Since $k_n = k_0\lambda^n$, the solution of Eq.~(\ref{eqS1}) can be written as 
\begin{equation}
u_n(k_0,t) = k_0^{-1}U_n(t),
\label{eqS3}
\end{equation} 
where $U_n$ satisfies the equation
\begin{equation}
\frac{\partial U_n}{\partial t}
=
i\left[
\lambda^{n+1}U_{n+2}U^*_{n+1}
-(1+c)\lambda^nU_{n+1}U^*_{n-1}
-c\lambda^{n-1}U_{n-1}U_{n-2}
\right],\quad n \ge 0,
\label{eqS4}
\end{equation} 
with the initial conditions
\begin{equation}
t = 0:\quad U_0 = i,\quad U_1 = 2,\quad U_n = 0,\quad n \ge 2.
\label{eqS5}
\end{equation} 
According to Eqs.~(\ref{eqB3}), (\ref{eqB4}) and (\ref{eqS3}), 
the Fourier transformed solution $u(k,t)$ of the continuous model is given by
\begin{equation}
u(k,t) = k^{-1/2}u_n(k_0,t) = k^{-1/2}k_0^{-1} U_n(t),\quad
k = k_n^{2/3} = k_0^{2/3} \sigma^n,
\label{eqS6}
\end{equation} 
with the reality condition $u(-k) = u^*(k)$ for negative $k$. 
Recall that $u(k) = 0$ for $k = 0$ due to the vanishing mean value condition, 
$\int u\,dx = 0$, assumed in Section~\ref{sec:Model}.  
The corresponding solution $u(x,t)$ in physical space is given by the 
inverse Fourier transform. 

\begin{figure}
\centering
\includegraphics[width = 0.9\textwidth]{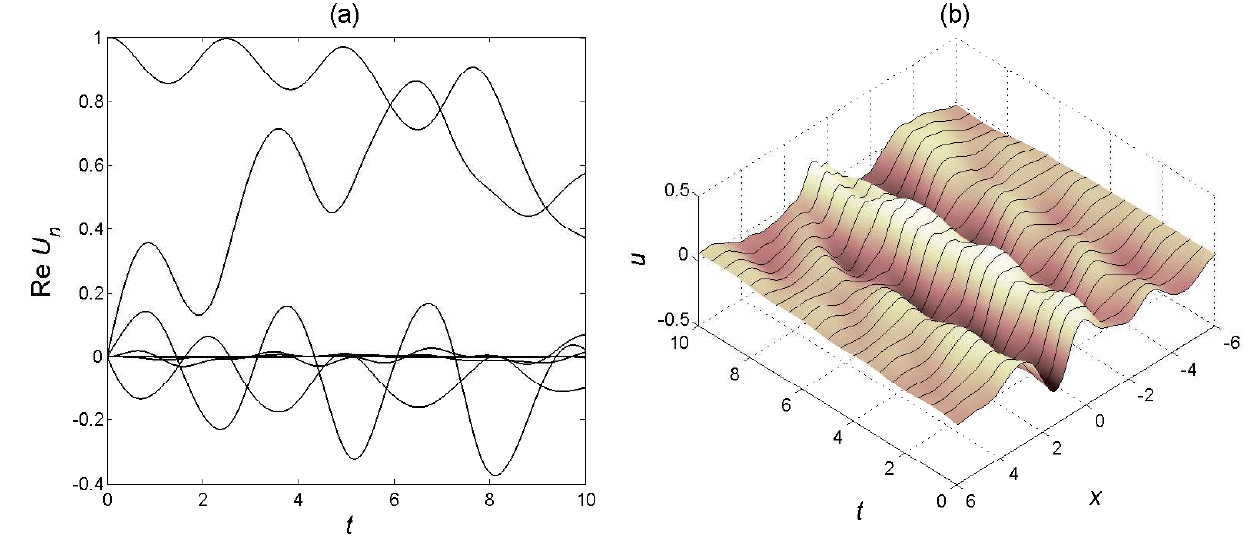}
\caption{Inviscid Sabra model with no forcing in the 2D regime ($c = \lambda^{-2}$). 
(a) Evolution of complex shell velocities $U_n(t)$ for initial conditions (\ref{eqS5}). 
Only real parts are shown. (b) Physical space representation $u(x,t)$ of the shell 
model solution $U_n(t)$.}
\label{fig4}
\end{figure}

The solutions can be found numerically with high accuracy by considering a finite 
number of shells, as it was done in Section~\ref{sec:DNvisc}. The numerical results 
are presented in Fig.~\ref{fig4}, showing the physical space 
representation of the single solution $U_n(t)$ for the Sabra model in 2D regime. 
This solution is smooth as expected due to regularity of the inviscid 2D Sabra 
model. 

Now let us consider the case $c = -\lambda^{-1} \approx -0.486$. In this case the 
conserved quantity (\ref{eqH}) is not sign definite and 
can be associated with the helicity. For this reason, Eq.~(\ref{eqS1}) with 
$c = -\lambda^{-1}$ is considered as the shell model for three-dimensional inviscid 
flow (3D Euler equations). For this Sabra model, there exists a unique local in time solution if 
the initial conditions have the finite norm $\sum_n k_n^2|u_n|^2 < \infty$, 
see~\cite{constantin2007regularity}. In general, the solution leads to a 
finite-time blowup~\cite{mailybaev2012} characterized by the infinite norm 
$\sum_n k_n^2|u_n|^2 \to \infty$ as $t \to t_b-0$. This blowup has the self-similar 
asymptotic form for large shells $n$ and $t \to t_b$. This form, up to system symmetries, 
is given by the expression $u_n(t) \to -ik_n^{-y}U(k_n^{1-y}(t-t_b))$, where $y$ and $U(t)$ 
are the universal real scaling exponent and real function depending only on the Sabra model 
parameter $c$~\cite{mailybaev2013blowup}.

\begin{figure}
\centering
\includegraphics[width = 0.99\textwidth]{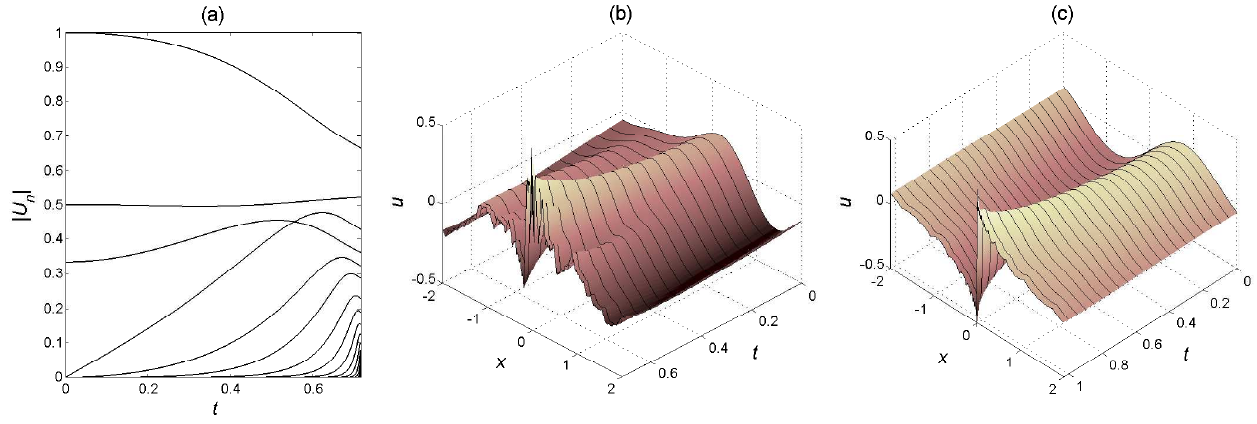}
\caption{Finite-time blowup in inviscid Sabra model with no forcing in the 
3D regime ($c = -\lambda^{-1}$). (a) Evolution of shell speed amplitudes $|U_n(t)|$ for 
complex initial conditions (\ref{eqS5}). 
(b) Physical space representation $u(x,t)$ of the shell model solution $U_n(t)$. 
(c) Physical space representation $u(x,t)$ of the purely imaginary 
solution $U_n(t)$. The last time in all the figures corresponds to the blowup point.}
\label{fig5}
\end{figure}

For numerical solution, we consider complex initial conditions (\ref{eqS2}) leading to the solution (\ref{eqS3}) of the system (\ref{eqS4}), (\ref{eqS5}) with $c = -\lambda^{-1}$. The solution $U_n(t)$ found numerically is presented in Fig.~\ref{fig5}(a). The blowup occurs at $t_b \approx 0.716$, and one can recognize the asymptotic self-similar form of this solution developing for large $n$ near the blowup time. The corresponding physical space solution $u(x,t)$ of the continuous model is obtained using Eq.~(\ref{eqS6}) with the inverse Fourier transform. The result is shown in Fig.~\ref{fig5}(b). One can observe the strongly nonlocal character of the blowup; this is not surprising due to nonlocality of the continuous model (\ref{eq1}), (\ref{eq2}). 

Despite the similarity of blowup structures for different shell models described 
by the renormalization method~\cite{dombre1998intermittency,mailybaev2012,mailybaev2012c}, the physical space 
representation of the blowup turns out to be very different; compare Fig.~\ref{fig5} 
with the blowup formation described in Figs.~\ref{fig2} and \ref{fig3}. 
For further insight, Fig.~\ref{fig5}(c) shows the blowup with purely imaginary initial 
conditions ($U_0(0) = 	-i/2$, $U_1(0) = -i$, $U_n(0) = 0$ for $n \ge 2$), 
which describe odd solutions $u(-x,t) = -u(x,t)$ of the continuous model. 
In this case, the blowup creates a discontinuity at $x = 0$ exactly at the blowup 
time. Note the difference from Fig.~\ref{fig3} (and also from the Burgers equation), where 
the discontinuity appears only after the blowup.  

\subsection{Viscous solutions}
\label{secSabra3}

In this section, we consider the viscous Sabra model (\ref{eqB7}) with the shells $n \ge 0$ 
($u_n = 0$ for $n < 0$). 
We assume vanishing initial conditions, 
the constant forcing with two nonzero elements $f_0 = ik_0^{-1}$ and $f_1 = (1+i)k_0^{-1}$, 
and viscous coefficients $\nu_n = \nu \lambda^{2n} = \nu (k_n/k_0)^2$. 
As we already mentioned, such choice of 
viscous terms corresponds to the continuous model with hyperviscosity. 

The shell model 
solution $u_n(k_0,t)$ can be written in the form (\ref{eqS3}),
where $U_n$ satisfy the equations
\begin{equation}
\frac{\partial U_n}{\partial t}
=
i\left[
\lambda^{n+1}U_{n+2}U^*_{n+1}
-(1+c)\lambda^nU_{n+1}U^*_{n-1}
-c\lambda^{n-1}U_{n-1}U_{n-2}
\right]+\nu \lambda^{2n} U_n + F_n
\label{eqT1}
\end{equation} 
for $n \ge 0$ 
with the nonzero forcing terms $F_0 = i$, $F_1 = 1+i$ and vanishing initial conditions.
Thus, a single solution $U_n(t)$ can be used to reconstruct the dynamics for all $k_0$. 
Then, the continuous model solution $u(k,t)$ is obtained using Eq.~(\ref{eqS6}) for positive $k$
and using the reality condition $u(-k) = u^*(k)$ for negative $k$. 
The corresponding solution $u(x,t)$ in physical space is given 
by the inverse Fourier transform. 

\begin{figure}
\centering
\includegraphics[width = 0.9\textwidth]{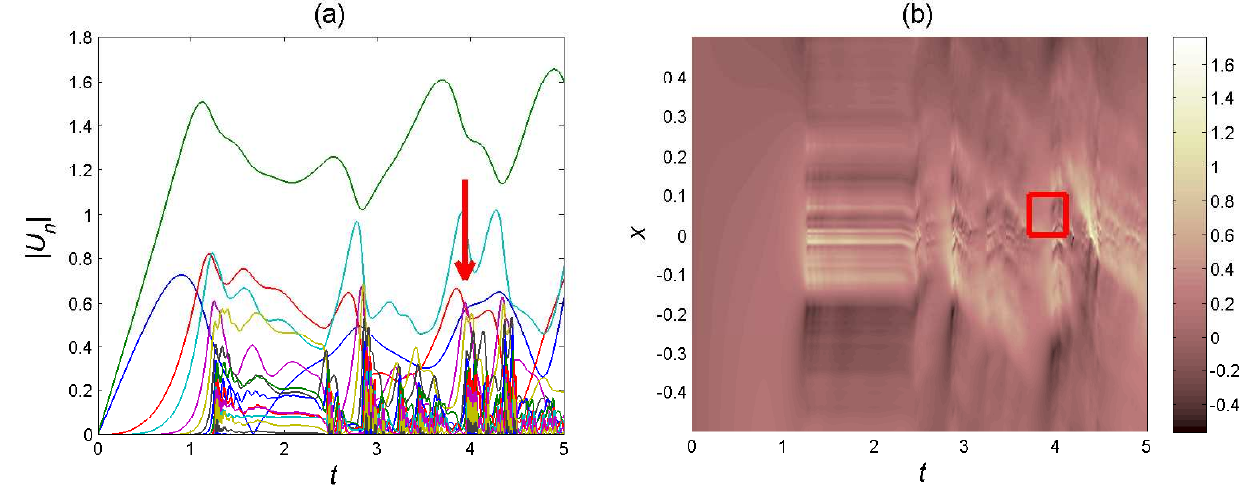}
\caption{(a) Dynamics of velocity amplitudes $|U_n(t)|$ for the Sabra model 
in 3D regime ($c = -\lambda^{-1}$)
for constant forcing and viscous coefficient $\nu = 10^{-5}$.
(b) Physical space representation $u(x,t)$ of the shell model solution. 
The red box in the right figure shows the part amplified in Fig.~\ref{fig7}; it corresponds to 
the formation of a turbulent burst pointed by the arrow in the left figure. }
\label{fig6}
\end{figure}

Figure~\ref{fig6}(a) presents the results of numerical simulations for the Sabra model (\ref{eqT1}). 
The singularity occurs near $t \approx 1.27$. 
This singularity corresponds to blowup in the inviscid system, while it is depleted 
at large shell numbers in the viscous model. After the transient period around 
$1.27 < t < 2.48$, the system enters into the intermittent turbulent regime. It is known 
that, in this regime, the inertial interval of scales (shells) is created, 
where the velocity moments 
scale as power-laws with the anomalous scaling exponents close to those in the Navier-Stokes 
turbulence~\cite{l1998improved,frisch1995turbulence}. The intermittent 
regime is characterized by a sequence 
of turbulent bursts. These bursts (also called instantons) have 
self-similar statistics for the Sabra model~\cite{mailybaev2012computation}, 
directly related to the anomalous scaling~\cite{mailybaev2013blowup}. 

\begin{figure}
\centering
\includegraphics[width = 0.9\textwidth]{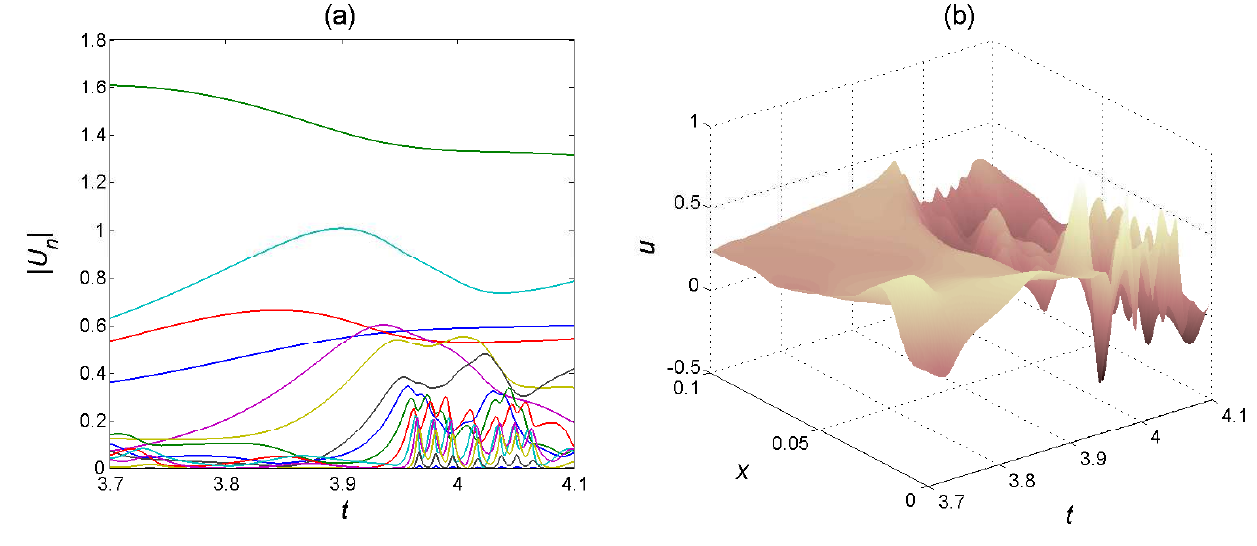}
\caption{Dynamics of velocity amplitudes $|U_n(t)|$ for the Sabra model 
in 3D regime ($c = -\lambda^{-1}$) and the corresponding 
physical space representation $u(x,t)$. The presented plots correspond to 
the specific turbulent burst indicated in Fig.~\ref{fig6}.}
\label{fig7}
\end{figure}

Figure~\ref{fig6}(b) presents the corresponding solution of the continuous model (\ref{eq1}) 
in the physical space $(x,t)$. One can see that the turbulent behavior is localized 
in the region around the origin, $x = 0$, 
due to localization of the constant forcing term $f(x)$. A sequence of 
turbulent bursts 
in physical space can be recognized, which correspond to the instantons 
in the Sabra shell model solution in Fig.~\ref{fig6}(a). We have chosen a specific instanton 
near $t \approx 4$ and showed its detailed structure for both shell amplitudes $|U_n|$ and 
physical space function $u(x,t)$ in Fig.~\ref{fig7}.

\section{Conclusion}

In this paper we constructed continuous hydrodynamic models, which reduce after the Fourier transformation to an infinite set of equivalent shell models. These continuous models share various properties of the Navier-Stokes equations like the scaling invariance, nonlocal quadratic term, energy conservation (for the inviscid system), etc. We also discuss their Hamiltonian representations. Such constructions are carried out for the dyadic (Desnyansky--Novikov) model with the intershell ratio $\lambda = 2^{3/2}$ and for the Sabra model of turbulence with $\lambda = \sqrt{2+\sqrt{5}} \approx 2.058$. Note that the values of $\lambda$ allowing the continuous representation are fixed and no limit like $\lambda \to 1$ is necessary. 

The constructed continuous models allow understanding various properties of the discrete shell models and provide their interpretation in physical space. This is especially pronounced for the dyadic shell model, where the asymptotic solution with Kolmogorov scaling, $u_n \propto k_n^{-1/3}$, is represented by a shock (discontinuity) for the induced continuous solution in physical space. Moreover, the finite-time blowup together with its viscous regularization in the continuous model follow the scenario similar to the Burgers equation. For the Sabra model, we provide the physical space representation for both blowup solution and intermittent turbulent dynamics. There is a drastic difference of these phenomena in physical space in comparison with the dyadic model. 

As a future research topic, it would be interesting to study the relation between the turbulent statistics of the Sabra model solutions with physical space properties of the corresponding intermittent continuous solutions. It is reasonable to expect the formation of inertial interval in the continuous model with the anomalous scaling for velocity moments~\cite{frisch1995turbulence} driven by similar dynamics of the Sabra model~\cite{l1998improved,biferale2003shell}. This relation is not expected to be simple. For example, the asymptotic solution of the dyadic shell model has the Kolmogorov scaling, 
while the shocks in the continuous representation should lead to the anomalous scaling similar to the Burgers equation turbulence~\cite{cardy2008non}. 

\section*{Acknowledgments} 
This work was supported by the CNPq (grant 305519/2012-3).

\bibliographystyle{plain}
\bibliography{refs}
 
\end{document}